\documentclass[conference]{IEEEtran}
\IEEEoverridecommandlockouts
\usepackage{cite}
\usepackage{amsmath,amssymb,amsfonts}
\usepackage{algorithmic}
\usepackage{hyperref}
\usepackage{graphicx}
\usepackage{textcomp}
\usepackage{xcolor}
\usepackage{multirow}
\usepackage{booktabs}
\usepackage{tabularx}

\newcolumntype{C}[1]{>{\centering\arraybackslash}p{#1}}

\def\BibTeX{{\rm B\kern-.05em{\sc i\kern-.025em b}\kern-.08em
    T\kern-.1667em\lower.7ex\hbox{E}\kern-.125emX}}
\begin{document}

\title{Towards Automated Classification of Attackers' TTPs by combining NLP with ML Techniques}

\author{\IEEEauthorblockN{Clemens Sauerwein and Alexander Pfohl}
\IEEEauthorblockA{\textit{Department of Computer Science} \\
\textit{University of Innsbruck}\\
Innsbruck, Austria \\
Firstname.Secondname@uibk.ac.at}
}

\maketitle

\begin{abstract}
The increasingly sophisticated and growing number of threat actors along with the sheer speed at which cyber attacks unfold, make timely identification of attacks imperative to an organisations' security. Consequently, persons responsible for security employ a large variety of information sources concerning emerging attacks, attackers' course of actions or indicators of compromise. However, a vast amount of the needed security information is available in unstructured textual form, which complicates the automated and timely extraction of attackers' Tactics, Techniques and Procedures (TTPs). In order to address this problem we systematically evaluate and compare different Natural Language Processing (NLP) and machine learning techniques used for security information extraction in research. Based on our investigations we propose a data processing pipeline that automatically classifies unstructured text according to attackers' tactics and techniques derived from a knowledge base of adversary tactics, techniques and procedures. 
\end{abstract}


\section{Introduction}
\label{sec:intro}
The increasing complexity, heterogeneity and globally distribution of modern IT systems combined with the increasing number of sophisticated cyber-attacks manifest serious threats harming an IT system’s security. Recent high-profile cyber security events (e.g. WannaCry~\cite{mohurle2017brief}, Meltdown~\cite{lipp2018meltdown} or Spectre~\cite{kocher2019spectre}) have shown that the spectrum of possible attacks steadily increases and that the time frame for organizations to react shrinks constantly~\cite{jang2014survey}. Accordingly, stakeholders' responsible for cyber security employ a large variety of security information sources to obtain information regarding attackers' tactics, techniques and procedures (TTPs)~\cite{sauerwein2019analysis}. These sources range from unstructured publicly available information sources~\cite{sauerwein2019analysis} (e.g. mailing lists, expert blogs, social media) to inter-organizational cyber threat intelligence sharing platforms~\cite{sauerwein2017threat,tounsi2018survey,wagner2016misp}. In this context, a tactic can be considered as a high-level description of an attacker's behavior, techniques provide a more detailed description of a behavior in the context of a tactic, and procedures an even low-level and detailed description in the context of a technique~\cite{johnson2016guide}. 

Although there are standards (e.g. STIX, OpenIoC,...) for describing TTPs, the needed information is usually available in unstructured textual form and cannot be automatically processed or analysed. However, the situation is exacerbated by the presence of a large amount of security data, which makes it difficult to capture TTPs in a timely manner through manual processes and makes security a big data problem~\cite{macdonald2012information}. Accordingly, former research suggests to implement the intelligence cycle~\cite{dempsey2013joint}, including a \textit{Processing \& Exploitation} step that translates security information into machine-readable format, to automatically extract TTPs from unstructured security data~\cite{bauer2020towards,de2020methodology}.  While there are some approaches that deal with the extraction of security information from unstructured text in general, it is unclear which approach is suitable for extracting TTPs from unstructured text. Furthermore, this challenge is highlighted by the fact that the OSINT goldmine of threat intelligence has not yet been comprehensively tapped and offers great potential for future research and practice~\cite{pastor2020not}. To address this research gap, we  seek to answer the  following research questions: (1) \textit{What are the current approaches to extract security information from unstructured text by using NLP and ML techniques?} (2) \textit{Which NLP and ML techniques are suitable to extract attackers' tactics, techniques and procedures from unstructured threat information?}

In order to address these research questions, we conducted a systematic literature study to identify how research combines
NLP and ML techniques to automatically extract security information from unstructured text. In a further step, we evaluated and compared the most commonly used techniques in order to identify those approaches which are most suitable to automatically extract attackers' TTPs. In doing so, we created a training set as input for these techniques based on the ATT\&CK~\cite{strom2018mitre} framework. Last but not least, we propose a processing pipeline which might be used to automatically extract attackers' tactics and techniques.
The remainder of this paper is structured as follows: Section~\ref{sec:rw} discusses related work regarding security information extraction. Section~\ref{sec:method} outlines the applied research methodology to identify, evaluate and compare NLP and corresponding ML techniques for the automated extraction of attackers' TTPs. Section~\ref{sec:results} provides an overview of the used NLP and ML techniques, evaluates them and proposes an information processing pipeline to automatically extract attackers' TTPs. Section~\ref{sec:discuss} discusses the key findings and limitations of the research at hand. Section~\ref{sec:conclusio} concludes the paper and provides outlook on future work.    
\section{Related work}
\label{sec:rw}
In the past several authors focused on the detection, extraction and analysis of security information from different data sources~\cite{rahman2020literature}. They primarily focused on the extraction of indicators of compromise~\cite{liao2016acing,zhao2020timiner}, vulnerabilities~\cite{sauerwein2018tweet,mulwad2011extracting,liu2012network,gawron2017automatic,han2017learning}, security events~\cite{ritter2015weakly}, exploits~\cite{sabottke2015vulnerability} and other security related information~\cite{joshi2013extracting,mittal2016cybertwitter,gharge2017integrated,vadapalli2018twitterosint,niakanlahiji2018natural,suh2016text,peng2018detecting}. In doing so, they rely  on sources like hacker forums~\cite{macdonald2015identifying,benjamin2015exploring,nunes2016darknet,samtani2016azsecure,deliu2017extracting,tavabi2018darkembed,williams2018incremental}, social media~\cite{sauerwein2018tweet,ritter2015weakly,sabottke2015vulnerability,mittal2016cybertwitter,gharge2017integrated,vadapalli2018twitterosint,niakanlahiji2018natural} and other sources~\cite{suh2016text,buldakova2014development,jones2015towards,li2018security,neil2018mining}. 

However, only a few contributions focused on the detection and extraction of TTPs in the last few years. For example, TTPDrill automatically extracts TTPs from unstructured cyber threat intelligence reports based on a manually created threat ontology~\cite{husari2017ttpdrill}. The same authors developed ActionMiner a framework that is able to extract a structured list of threat actions from unstructured threat reports~\cite{husari2018using}. The MITRE corporation developed the Threat Report ATT\&CK Mapper (TRAM)\footnote{\url{https://github.com/mitre-attack/tram} (Accessed: August 2nd, 2021)}, a web based tool for extracting adversary techniques from cyber threat intelligence reports in order to map the extracted information to ATT\&CK~\cite{strom2018mitre}. Moreover, the rcATT\footnote{\url{https://github.com/vlegoy/rcATT} (Accessed: August 2nd, 2021)} tool bases on an old version of the ATT\&CK framework and automatically extracts tactics and techniques from unstructured threat reports~\cite{legoy2020automated}. Last but not least,~\cite{rahman2020literature} provides a literature review to understand the source, purpose, and approaches for mining cyberthreat intelligence from unstructured text.

A variety of approaches and data sources have been used to extract security information from structured and unstructured texts in the past. Most of them focus on one specific method for each task (e.g. using TF-IDF for vectorization), but do not consider other methods and compare them among each other. Our research addresses this gap by evaluating and comparing different approaches that combine NLP and ML techniques to extract relevant security information. Finally, based on these evaluations we propose a processing pipeline for extracting TTPs from unstructured texts.

\section{Applied research methodology}
\label{sec:method}
In order to investigate the used NLP and ML techniques to extract TTPs from unstructured texts, we  we conducted a systematic literature study (see Section~\ref{subsec:slr}) to identify different approaches combining NLP and ML techniques followed by an evaluation of the identified techniques (see Section~\ref{subsec:eval}) by using an adversary training data set generated from the ATT\&CK~\cite{strom2018mitre} knowledge base. Based on our investigations we propose a data processing pipeline for the classification of TTPs from unstructured text.

\subsection{Systematic Literature Study}
\label{subsec:slr}
The goal of this literature study was to systematize current approaches that combine NLP and ML techniques to extract security-related information form unstructured texts. To achieve this goal we conducted a systematic literature study based on the methodologies by~\cite{kitchenham2004procedures} and~\cite{wohlin2014guidelines} between January 2021 and February 2021. In doing so, we applied the following procedure: \textit{definition of search strategy}, \textit{initial search}, \textit{paper selection}, \textit{snowballing} and \textit{data extraction}. In order to guarantee repeatability and traceability of the research methodology, a review protocol, including search strategy, search results, selection criteria, selection procedure and data extraction was developed. In the course of the \textit{definition of the search strategy} and with respect to our research objective and questions, we defined the following search term for our \textit{initial search}: \textit{("information security" OR "vulnerability" OR "vulnerabilities" OR "cyber security" OR "cybersecurity" OR "threat intelligence" OR "threat" OR "threats" OR "cyber threat intelligence" ) AND ("data mining" OR "information extraction" OR "extracting" OR "extraction OR "natural language processing" OR "NLP" OR "machine learning")}. We used the databases ACM Digital  Library,  AIS  Electronic  Library,  EBSCOhost, IEEE Xplore Digital Library, Springer Link, Taylor \& Francis Online and Wiley Online Library. Only papers published between 2009 and 2020 were considered. Our \textit{initial search} delivered 5307 papers in total. In the next step we performed the \textit{paper selection}. In doing so, we (i) eliminated all duplicates and excluded all papers that (ii) were not peer-reviewed or are grey or white literature,  (iii) were not available in full text and (iv) have not used a combination of NLP and ML techniques. By reading the title, abstracts, and keywords and by skim reading of full texts of all papers, we (v) assessed their relevance for answering our research questions (see Section~\ref{sec:intro}). The \textit{paper selection} resulted in a set of 20 papers. To avoid overlooking relevant papers, we performed one iteration of backward \textit{snowballing} based on~\cite{wohlin2014guidelines}. By reviewing the citations of our 20 selected papers on Google Scholar with respect to our \textit{paper selection} procedure, we identified additional 6 relevant papers. This procedure resulted in a final set of 26 papers. Finally, we performed \textit{data extraction} on the final set of 26 papers based on a concept matrix. This matrix included the following classification categories: (i) used NLP techniques and (ii) ML techniques.

\newcolumntype{C}[1]{>{\centering\arraybackslash}p{#1}}
\begin{table*}[!t]
\centering
\caption{Overview of NLP and ML Techniques used in relevant works to extract security-related information form unstructured texts}
\label{tab:slrresults}
\begin{tabular}{@{}llC{0cm}C{0cm}C{0cm}C{0cm}C{0cm}C{0cm}C{0cm}C{0cm}C{0cm}C{0cm}C{0cm}C{0cm}C{0cm}C{0cm}C{0cm}C{0cm}C{0cm}C{0cm}C{0cm}C{0cm}C{0cm}C{0cm}C{0cm}C{0cm}C{0cm}C{0cm}C{0cm}c@{}}
\toprule
                      &    & \rotatebox{90}{\cite{husari2017ttpdrill}} & \rotatebox{90}{\cite{purba2020word}}  & \rotatebox{90}{\cite{xun2020aiti}} & \rotatebox{90}{\cite{niakanlahiji2018natural}} & \rotatebox{90}{\cite{ghazi2018supervised}} & \rotatebox{90}{\cite{liao2016acing}}  & \rotatebox{90}{\cite{zhao2020timiner}}  & \rotatebox{90}{\cite{mulwad2011extracting}} & \rotatebox{90}{\cite{liu2012network}}  & \rotatebox{90}{\cite{joshi2013extracting}} & \rotatebox{90}{\cite{suh2016text}} & \rotatebox{90}{\cite{han2017learning}}   & \rotatebox{90}{\cite{gawron2017automatic}}  & \rotatebox{90}{\cite{peng2018detecting}}   & \rotatebox{90}{\cite{iorga2020early}} & \rotatebox{90}{\cite{satyapanich2020casie}}  & \rotatebox{90}{\cite{benjamin2015exploring}}  &  
                      \rotatebox{90}{\cite{samtani2016azsecure}} & \rotatebox{90}{\cite{deliu2017extracting}}  & \rotatebox{90}{\cite{tavabi2018darkembed}}   & \rotatebox{90}{\cite{ritter2015weakly}}   & \rotatebox{90}{\cite{mittal2016cybertwitter}}   & \rotatebox{90}{\cite{vadapalli2018twitterosint}}    & \rotatebox{90}{\cite{niakanlahiji2019iocminer}}      & \rotatebox{90}{\cite{jones2015towards}}  & 
                      \rotatebox{90}{\cite{li2018security}}   &   \\ \midrule
\multirow{8}{*}{\rotatebox{90}{\textbf{NLP}}}                 
& \textbf{Tokenization}   &    & 1   & 1   & 1   & 1   & 1   & 1   &    & 1   &    &    & 1   &    & 1   &    & 1   &        &    & 1   & 1       &        &        & 1   & 1       & 1       & 1    &    \\
& \textbf{POS Tagging}    & 1   & 1   &    & 1   & 1   & 1   & 1   &    & 1   &    &    & 1   &    &    &    & 1   &       &    & 1   & 1       &        &        & 1   & 1       & 1       & 1    &   \\
& \textbf{NER}            &    &    &    &    & 1   & 1   & 1   & 1   &    & 1   & 1   &    &    &    &    & 1    & 1       &    & 1   &        & 1   & 1       & 1   & 1       &        & 1      &\\
& \textbf{Stemming}       &    & 1   &    & 1   &    &    &    &    & 1   &    &    &    & 1   &    &    &    &       &    &    &        &        &        &    & 1       &        &       & \\
& \textbf{Lemmatization}       &    & 1   &    &    &    &    &    &    &    &    &    &    &    &    &    & 1   &        &    &    &        &       &        &    &        &        &       & \\
& \textbf{Bag-of-Words}   &    &    &    &    &    &    &    & 1   &    &    &    &    & 1   &    &    &    & 1      & 1   & 1   &        &        &        &    & 1       &        &     &    \\
& \textbf{TF-IDF}         & 1   &    &    & 1   &    &    &    &    & 1   &    & 1   &    &    &    & 1   &    & 1   &       &    &        &       &        &    &        &        & 1   &    \\
& \textbf{Word Embedding} &    &    & 1   &    &    &    & 1   &    &    &    &    & 1   & 1   &    &    &    &    &       & 1   & 1       &        &        &    &        &        &    &    \\ \midrule
\multirow{3}{*}{\rotatebox{90}{\textbf{ML}}} 

& \textbf{SVM}            & 1   &    &    &    &    & 1   &    & 1   & 1   &    & 1   & 1   &    & 1   & 1   &    &       & 1   & 1   & 1       &      &       &    &        &        & 1   &    \\
& \textbf{NB}             &    &    &    &    &    &    &    &    & 1   &    & 1   &    & 1   & 1   &    &    &    &     &    &        &        &        &    &        &        &    &    \\
& \textbf{Other}          &    &  1  & 1   & 1   &  1  &    & 1  &    &    &  1  &    &    &    &    &    &  1  & 1    &        &    &       & 1        &  1  & 1   & 1       & 1       &    &    \\ \bottomrule
\end{tabular}
\end{table*}
\subsection{Evaluation of NLP and ML Techniques}

Based on the results of the systematic literature study we selected the most common NLP and ML techniques to evaluate their use for the extraction of TTPs from unstructured texts. In doing so, different processing pipelines have been built based on all possible combinations of the identified NLP and most widely used ML techniques (see Section~\ref{subsec:pipeline}).  In order to implement the NLP techniques we used NLTK\footnote{\url{https://www.nltk.org/} (Accessed: August 2nd, 2021)} and gensim\footnote{\url{https://pypi.org/project/gensim/} (Accessed: August 2nd, 2021)}. For the the ML classification task we used scikit-learn\footnote{\url{https://scikit-learn.org/} (Accessed: August 2nd, 2021)}. A training set was automatically created based on the ATT\&CK framework~\cite{strom2018mitre} (see Section~\ref{subsec:pipeline}). In order to evaluate the whole processing pipeline we used performance measures and five-fold cross-validation (see Section~\ref{subsec:eval}. In doing so we used measures like micro- and macro-averaging scores of precision, recall and $F_{0.5}$-score as well as Hamming loss of the different models.

\section{Results}
\label{sec:results}
In this section we provide a comprehensive overview of approaches combining NLP and ML \textit{techniques to extract security information from unstructured text} (see Section~\ref{subsec:overview}), present a processing pipeline combining the most common NLP and ML techniques identified in literature (see Section~\ref{subsec:pipeline}) and its evaluation (see Section~\ref{subsec:eval}).
\subsection{Techniques to extract security information from unstructured text}
\label{subsec:overview}
Firstly, our analysis reveals that a plethora of NLP approaches were used in the identified literature. As depicted in table~\ref{tab:slrresults} the following NLP techniques were used in the analyzed papers: \textit{Tokenization} (16), \textit{Part-of-Speech (POS) Tagging} (15), \textit{Named-Entity-Recognition (NER)} (14), \textit{Stemming} (5) and \textit{Lemmatization} (2). Moreover, we identified  additional approaches like \textit{lower-casing}~\cite{gawron2017automatic}, customized extraction algorithms (e.g.~\cite{liao2016acing}) or the creation of a context-free grammar~\cite{niakanlahiji2018natural}. These results indicate that NLP approaches are used frequently, yet not all steps are reported in the papers. For example \textit{POS} by implication is always preceded by \textit{Tokenization}, yet this stage wasn't mentioned in~\cite{husari2017ttpdrill}.

Secondly, during our analysis, we grouped ML approaches according to the frequency of their use in the identified literature. Accordingly, we elicted the following categories: \textit{Support Vector Machine (SVM)} (12), \textit{Naive Bayes (NB)} (4) and others. Among the latter were approaches like \textit{Neural Network (NN)} based Approaches, \textit{Conditional Random Fields (CRF)}, \textit{Logistic Regression} and \textit{Random Forest}. The reason why SVMs are the first choice might be trace back to their easy-to-use nature and aptitude for classification problems. 
\subsection{Processing Pipeline to extract attackers' TTPs}
\label{subsec:pipeline}
\begin{figure}[h]
    \centering
    \includegraphics[width=\columnwidth]{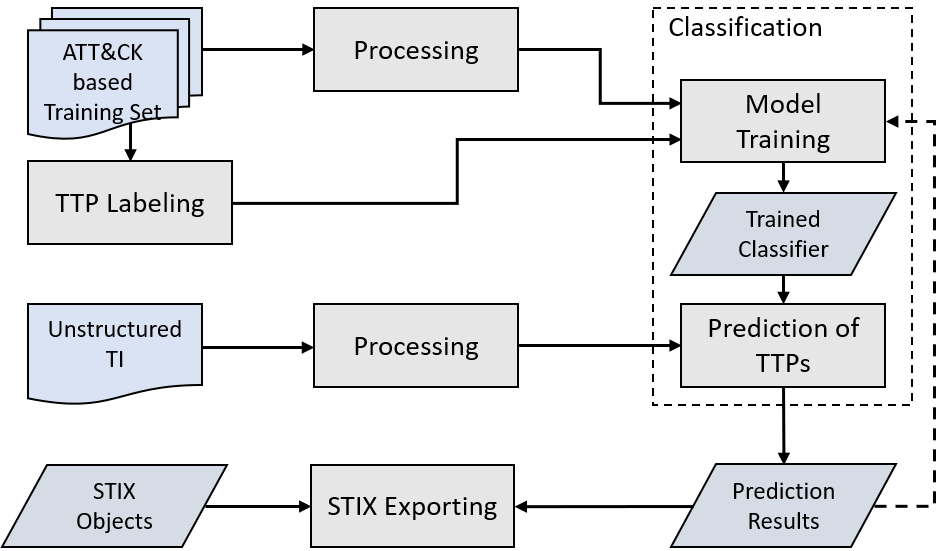}
    \caption{Processing pipeline to extract attackers' tactics and techniques}
    \label{fig:overview}
\end{figure}

Based on the identified NLP and ML techniques, we defined an evaluation processing pipeline to extract attackers' tactics and techniques from unstructured text and transform them to a standardized data format. It is worth mentioning that we did not consider attackers' procedures since our training set doesn't provide labeled data on them. As depicted in Figure~\ref{fig:overview} our evaluation processing pipeline consists of four processes - \textit{TTP Labeling}, \textit{Processing}, \textit{Classification} and \textit{STIX Exporting}. 

As mentioned in Section~\ref{subsec:overview}, all identified approaches used a form of supervised learning. Therefore, we implemented a \textit{TTP Labeling} process that generates a labeled data set as input for the \textit{Classification} and final cross-validation. However, our analyzed papers and related work indicated a lack of publicly available training set for classifying TTPs. Therefore, we decided to rely on the ATT\&CK~\cite{strom2018mitre} framework to automatically create the needed training set. Within the ATT\&CK Knowledge base, reports and threat descriptions are already part of the descriptions of specific tactics and techniques. Accordingly, their labels are implicitly given and manual labeling is not needed. Therefore we extracted the data from the ATT\&CK GitHub Repository, extracted the content of each document, filtered them and stored them for further processing. In doing so we processed 534 documents, including 417 HTML and 117 PDF documents from 139 different sources.

The next step includes the \textit{Pre-processing}, \textit{Normalization}, \textit{Vectorization} and \textit{Transformation} of the  security and \textit{ATT\&CK based training set}. The aim of this \textit{Processing} step is the preparation of the data for \textit{Classification}.  In order to use appropriate techniques, the most common NLP techniques in the analyzed literature (cf. Section~\ref{subsec:slr}) were used. For the \textit{Pre-processing} we use \textit{Tokenization}, \textit{POS Tagging} and \textit{IoC Replacement}. Accordingly, after each IoC in the paragraphs was replaced by the appropriate natural language term, tokenization was performed and each token was tagged by its part-of-speech (POS). This process was performed for each paragraph in the security and training data set. On the obtained results \textit{Normalization} was applied, which can be achieved by \textit{Stemming} or \textit{Lemmatization}. In order to improve \textit{Normalization}, additional functionalities, like lower-casing, stopwords removal, single-charcater and non-word token removal were implemented. After \textit{Normalization} we performed \textit{Vectorization} by implementing several alternative \textit{Vectorization} techniques. During our systematic literature review several \textit{Vectorization} techniques were identified, namely \textit{Bag-of-word} features, \textit{TF-IDF} and \textit{Word Embeddings}. Since \textit{Bag-of-words} usually refer to word frequencies, but none of the analyzed papers mentioned the used technique explicitly, we decided to implement \textit{One-hot-endocing}, \textit{Term Frequency} and \textit{TF-IDF}. Moreover, since \textit{Word Embeddings} were discussed quite often in the literature we decided to implement \textit{Doc2Vec} as well. Accordingly, we implemented four possibilities to perform \textit{Vectorization}.

As shown by our systematic literature study (cf Section~\ref{subsec:slr}, for \textit{Classification} current research primarily focus on \textit{Support Vector Machines (SVM)} and \textit{Naive Bayes (NB)}. Therefore, we decided to include all of the \textit{SVM} and \textit{NB} classifiers provided by the scikit-learn library considering that they are compatible with our previous \textit{Vectorization} techniques and the multi-label classification problem. In order to fulfill this, we implemented a \textit{Transformation} process for the \textit{Vectorization} results. In doing so we implemented the \textit{Transformation} to the following representations: \textit{Binary Relevance}, \textit{Classifier Chain} and \textit{Distributed Representation}. For each of the representations, we used the following classifiers for evaluation: \textit{MutlinominialNB}, \textit{ComplementNB}, \textit{GaussianNB}, \textit{SVC}, \textit{LinearSVC} and \textit{SGDClassifier}. In doing so, the \textit{Classification} consists of \textit{Model training} based on the vectorized and labeled ATT\&CK training set and the \textit{Prediction of TTPs} from unstructured security data based on the \textit{Trained Classifier}. 

Finally, we transformed the \textit{Prediction Results} to STIX format using the \textit{STIX Exporting} process. In doing so, each of the resulting predictions was annotated with respective labels for tactics and techniques. These labels were then used to query the ATT\&CK Repository for the corresponding \textit{STIX Objects} to those labels. 

\subsection{Proposed processing pipeline to extract attackers' tactics}
\label{subsec:eval}
In order to find the most suitable processing pipeline for the extraction of tactics and techniques, the processing and classification was closely scrutinized. In the course of this, all possible combinations of processing and classifications techniques were evaluated and compared. The following expression describes all possible combinations that were considered for the evaluation: \textit{Preprocessing AND (Steamming XOR Lemmatization) AND (One-hot encoding XOR Term-Frequency XOR TF-IDF OR Doc2Vec) AND (Binary Relevance XOR Classifier-Chain XOR Distributed-Representations) AND (MutlinominialNB XOR ComplementNB XOR GaussianNB XOR SVC XOR LinearSVC XOR SGDClassifier)}. In this context it is worth mentioning that all combinations were possible, except \textit{MultinominialNB} and \textit{ComplementNB} were not compatible with \textit{Doc2Vec}. Overall, these combinations resulted in 156 different pipelines for evaluation. As described in Section~\ref{subsec:eval}, we used performance measures and five-fold cross-validation to evaluate these pipelines and find the best performing combination of NLP and ML techniques for the extraction of tactics and techniques. 

Finally, our evaluations showed that \textit{Lemmatization}, \textit{One-hot encoding} combined with \textit{Binary Relevance} for tactics and techniques classification are best suited for the proposed processing pipeline. For the classification of tactics it turned out that \textit{SGDClassifier} performed best, with a prescision of 71.09\%, recall of 62.55\%, $F_{0.5}$ score of 69.06\% and hamming-loss of 26.70\%. For the classification of techniques the most appropriate classifier was \textit{LinearSVC} with a precision of 53.43\%, recall of 14.88\%, $F_{0.5}$ of 33.68\% and hamming-loss of 5.55\%. 

\section{Discussion}
\label{sec:discuss}
In this section we discuss our key findings (see Section~\ref{subsec:keyfindings}) and limitations of our research (see Section~\ref{subsec:limiations}).
\subsection{Key Findings}
\label{subsec:keyfindings}
Our research has shown that the most powerful processing pipeline that combines NLP and ML techniques to extract attackers' techniques and tactics from unstructured text uses the following techniques:  (1) \textit{Tokenization}, (2) \textit{POS Tagging}, (3) \textit{IoC Replacement}, (4) \textit{Lemmatization}, (5) \textit{One-hot encoding Vectorization}, (6) \textit{Binary Relevance Transformation} and (7) \textit{Support Vector Machine Classification}. For the latter it turned out that for tactics the SGDClassifier and for techniques the LinearSVC performed best. Unfortunately, it was not possible to classify attackers' procedures due to a lack of corresponding training data. Moreover the classification of attackers' tactics and techniques was complicated by an imbalanced training set. However, these issues might be traced back to the fact that we used the ATT\&CK framework as basis for our automatically generated training set. Nonetheless, the lack of a publicly available training data set and the need for publicly available labeled and balanced training data sets was highlighted by other security researchers as well~\cite{rahman2020literature}.

In addition, during the evaluation of our processing pipelines, it became apparent that there is no real baseline or publicly available source code of comparable approaches. Therefore, the evaluation of our 156 different processing pipelines was complicated, and we decided to use the dummy classifier of the skicit-learn\footnote{\url{https://scikit-learn.org/} (Accessed: August 2nd, 2021)} as baseline. This baseline worked well for classifying tactics, but unfortunately was not able to classify techniques properly. The latter might be due to the already discussed problem of an imbalanced training set. 

Finally, we tried to compare our approach with related work. Unfortunately, as mentioned above, this was only possible to a limited extent due to the lack of a general baseline and different training data sets. For example, a comparison of the TRAM tool\footnote{\url{https://github.com/mitre-attack/tram} (Accessed: August 2nd, 2021)} with our approach showed that they only focus in the classification of techniques based in the ATT\&CK framework. In addition, we analyzed at the rcaTT tool~\cite{legoy2020automated}, which also classifies tactics and techniques. Unfortunately, it was very difficult to compare our approach the rcaTT tool, as it builds on an older version of ATT\&CK and integrates a lot of noise through their automatic method of generating the training data set by considering external references as well. These comparative trials and the problems associated with them, once again highlight the need for publicly available labeled training data sets and baselines.

\subsection{Limitations}
\label{subsec:limiations}
The research at hand might be limited by a (i) \textit{selection bias} and (ii) \textit{wrong classification of relevant publications} dealing with NLP and ML techniques used for security information extraction. Moreover, it might be (iii) \textit{limited training set} primarily focusing on tactics and techniques based on the ATT\&CK framework and a (iv) \textit{missing baseline} to compare the performance of ML techniques for TTP classification. In order to counteract (i) and guarantee a complete list of relevant papers focusing in NLP and ML techniques we conducted a systematic literature study based on~\cite{kitchenham2004procedures} combined with the snowballing method~\cite{wohlin2014guidelines}. In order to counteract (ii) a concept matrix was created, as described in~\cite{kitchenham2004procedures}, to systematically extract information on the used NLP and ML techniques and a review protocol was kept. Two of the authors of this publication did the classification of each paper. In a final step the concept matrices and the review protocols were compared. If a conflict occurred, consensus based on a re-reading of the publication and a subsequent discussion was reached. Limitation (iii) was deliberately chosen because the automatic generation of a labeled training set based on the ATT\&CK framework was the only efficient solution and the generation of a manual labeled data set was too time-consuming. In order to counteract (iv) we compared our processing pipeline to related work. A detailed comparison was made difficult because comparable contributions only made their evaluation data available to a limited extent. 

\section{Conclusion \& Outlook}
\label{sec:conclusio}
In this paper we investigated, how current NLP and ML approaches can be combined to extract attackers' TTPs from unstructured text. Therefore, we analyzed the state of the art regarding the combination of NLP with ML techniques to extract security-related information from unstructured text. We evaluated and compared the different techniques described in 26 publications. In doing so we evaluated all possible combinations of  the identified NLP and ML techniques with 156 processing pipelines and an automatically generated training set based on the ATT\&CK framework. It turned out that a combination of \textit{Tokenization}, \textit{POS Tagging}, \textit{IoC Replacement}, \textit{Lemmatization}, \textit{One-hot encoding}, \textit{Binary Relevance} and \textit{Support Vector Machine} performed best for the classification of techniques and tactics. However, due to a lack of training date we were not able to classify attackers' procedures. Finally, our investigations highlighted the need for publicly available labeled training set and a baseline for evaluation. Future work will include a comprehensive evaluation of further machine learning techniques and the creation of a publicly available training data set to classify attackers' TTPs. Moreover, we will focus on a processing pipeline and labeled training data set to extract attacker's procedures.
\bibliographystyle{ieeetr}
\bibliography{bibliography}
\end{document}